\documentclass{iau} 
\usepackage{graphicx}
\usepackage{subfig}

\title[] %% give here short title %%
{The stellar mass distribution of \\S$^{4}$G disk galaxies}

\author[]   %% give here short author list %%
{Sim\'on D\'iaz-Garc\'ia$^1$, Heikki Salo$^1$ \& Eija Laurikainen$^{1}$}
 
\affiliation{$^1$Astronomy Research Group, University of Oulu, \\ FI-90014 Finland,
\\ email: {\tt simon.diazgarcia@oulu.fi} \\[\affilskip]}

\pubyear{2015}
\volume{321}  %% insert here IAU Symposium No.
\jname{Formation and evolution of galaxy outskirts}
\editors{A. Gil de Paz, J. Knapen \& J. Lee, eds.}
\begin{document}

\maketitle

\begin{abstract}

We use 3.6 $\mu$m imaging from the S$^{4}$G survey to characterize the typical stellar density profiles ($\Sigma_{\ast}$) and bars as a function of fundamental galaxy parameters (e.g. the total stellar mass $M_{\ast}$), 
providing observational constraints for galaxy simulation models to be compared with.
We rescale galaxy images to a common frame determined by the size in physical units, by their disk scalelength, or by their bar size and orientation. 
We stack the resized images to obtain statistically representative average stellar disks and bars. 
For a given $M_{\ast}$ bin ($\ge 10^{9}M_{\odot}$), 
we find a significant difference in the stellar density profiles of barred and non-barred systems that gives evidence for bar-induced secular evolution of disk galaxies: 
(i) disks in barred galaxies show larger scalelengths and fainter extrapolated central surface brightnesses, 
(ii) the mean surface brightness profiles of barred and non-barred galaxies intersect each other slightly beyond the mean bar length, most likely at the bar corotation,
and (iii) the central mass concentration of barred galaxies is larger (by almost a factor 2 when $T<5$) than in their non-barred counterparts. 
We also show that early- and intermediate-type spirals ($0 \le T < 5$) host intrinsically narrower bars than the later types and S0s, whose bars are oval-shaped. 
We show a clear correlation between galaxy family and bar ellipticity. 


\end{abstract}

\firstsection

\section{Introduction}

% In the $\Lambda$CDM model, galaxies are formed in the center of dark matter halos when the gas cools down and condenses (e.g. \cite[White $\&$ Rees 1978]{White78}, \cite[Fall $\&$ Efstathiou 1980]{Fall80}). 
Approximately two-thirds of the galaxies in the local universe are barred (e.g. \cite[Knapen et al. 2000]{Knapen2000}, \cite[Laurikainen et al. 2004]{Laurikainen2004}). 
Simulation models predict that bars participate in the redistribution of stars and gas inside the galactic disk (e.g. \cite[Athanassoula 2013]{Athanassoula13}) by pushing them outwards (inwards) beyond (within) corotation. 
Bars are expected to increase the disk size (e.g., \cite[Hohl 1971]{}; \cite[Athanassoula $\&$ Misiriotis 2002]{}; \cite[Debattista et al. 2006]{}; \cite[Minchev et al. 2011]{}; \cite[Athanassoula et al. 2012]{}) 
and the central mass concentration after the bar-funneled cold gas is turned into stars (e.g., \cite[Athanassoula 1992]{}; \cite[Wada $\&$ Habe 1992]{}). 
Using 2-D decompositions, \cite[S\'anchez-Janssen $\&$ Gadotti (2013)]{} found observational evidence for the bar-driven disk evolution.
%
%-------------------------------------------------------------
%
\section{1-D average stellar density profiles}
%
%-------------------------------------------------------------
%
\begin{figure}[t]
% \vspace*{-2.0 cm}
\begin{center}
% \begin{tabular}{c c c c}
\subfloat{
\includegraphics[width=1.9in]{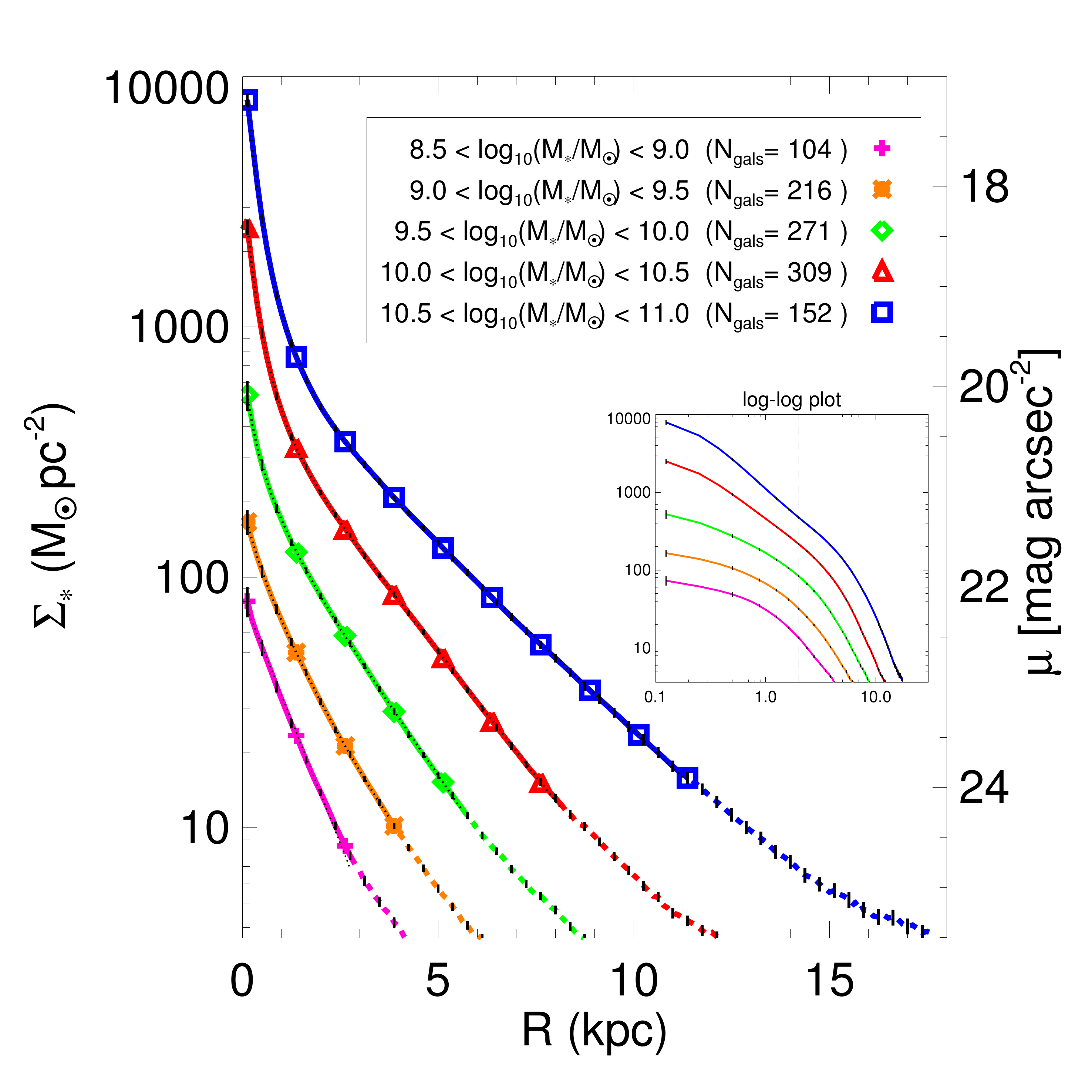}}
\subfloat{
\includegraphics[width=1.9in]{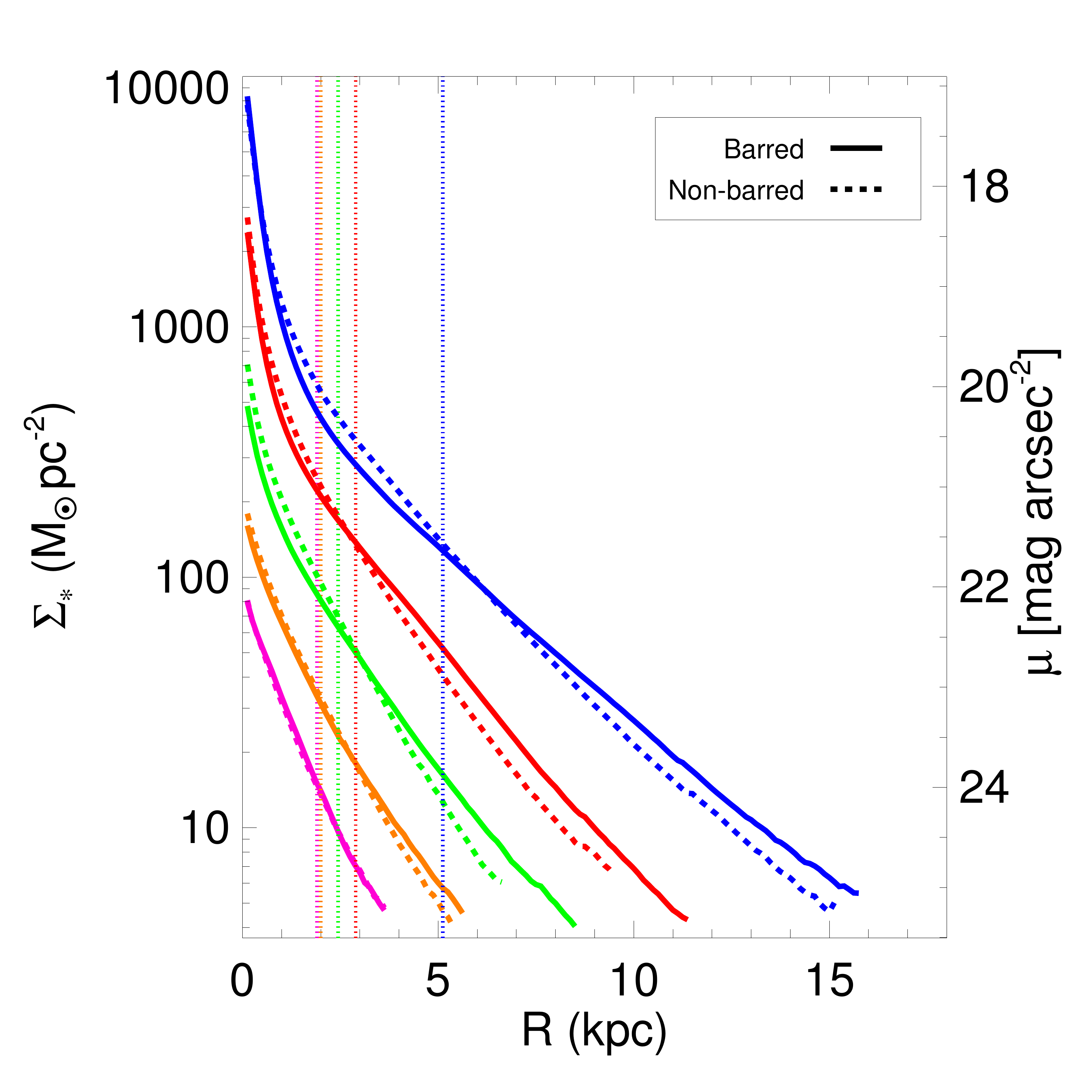}}\\
% \vspace*{-0.5 cm}
\vspace*{-0.75 cm}
\subfloat{
\includegraphics[width=1.9in]{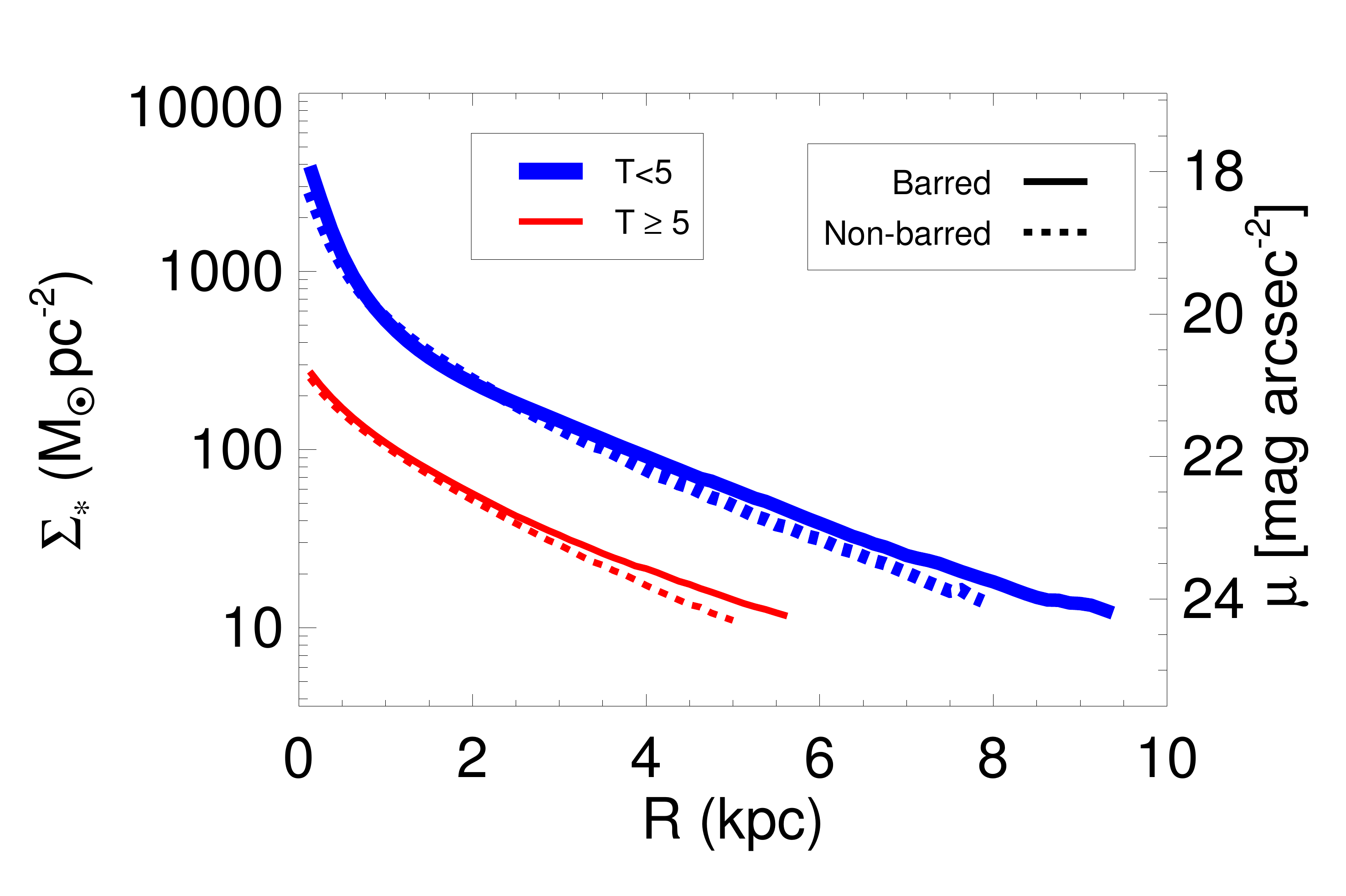}}
\subfloat{
\includegraphics[width=1.9in]{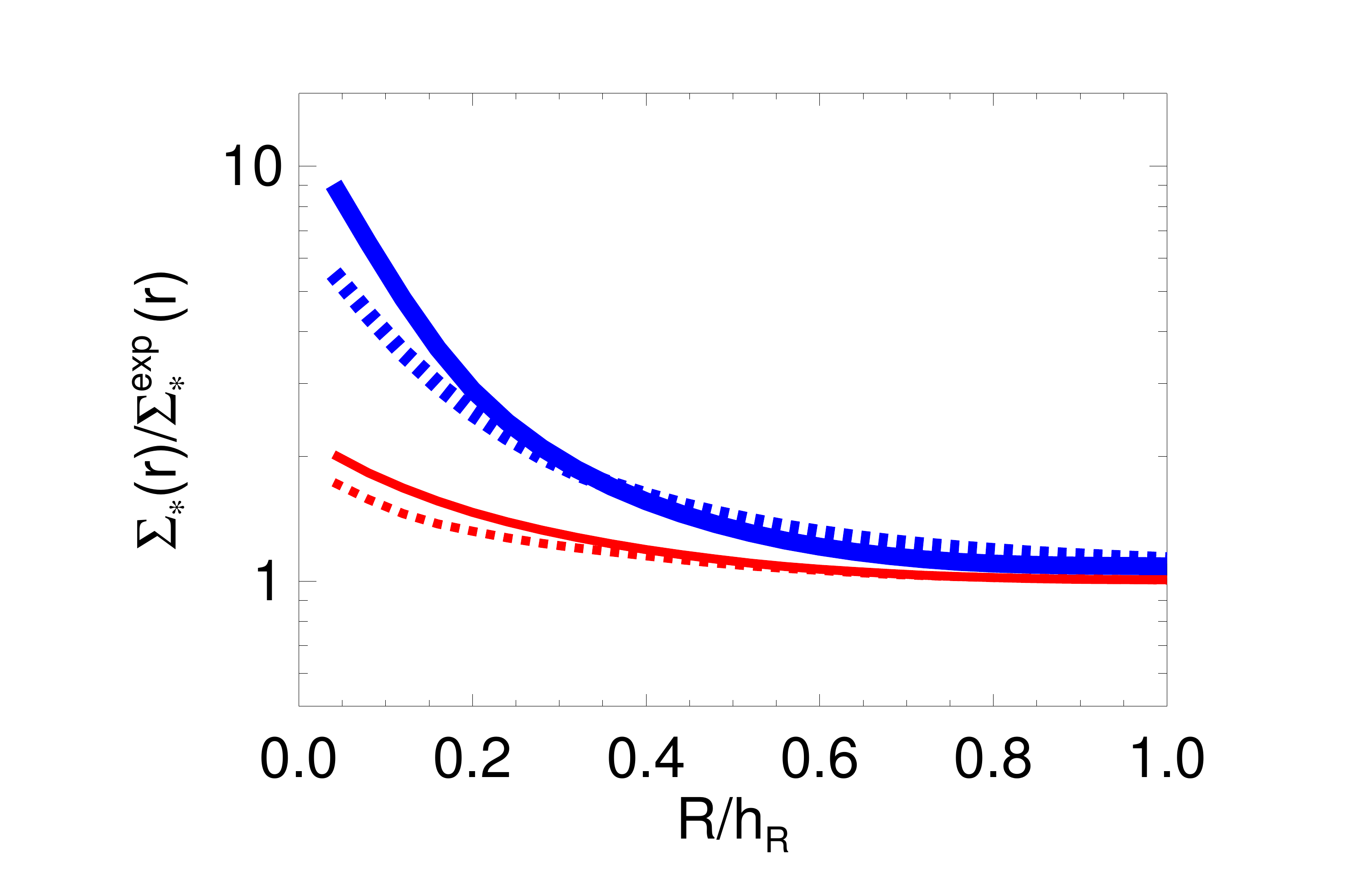}}\\
% \end{tabular}
% \vspace*{-0.32 cm}
\vspace*{-0.4 cm}
\caption{
\emph{Upper panels:} 
Average surface brightness profiles as a function of the total stellar mass (left). 
The solid (dashed) lines trace $\Sigma_{\ast}$ within the radial range with a $100\%$ ($75\%$) sample coverage. 
In the right panel we show the $M_{\ast}$-binned mean $\Sigma_{\ast}$ for barred (solid lines) and non-barred (dashed lines) systems ($90\%$ coverage). 
The vertical dotted lines indicate the mean bar size (measured in \cite[Herrera-Endoqui et al. 2015]{HE2015}) of the barred galaxies in each of the $M_{\ast}$-bins. 
\emph{Lower panels:} 
Mean $\Sigma_{\ast}$ profiles of barred (solid line) and non-barred galaxies (dashed lines), splitting the sample into early- (thick blue lines) and late-type (thin red lines) systems (left). 
With the same binning, in the right panel we show the deviation from an exponential disk ($\Sigma_{\ast}^{\rm exp}$) within the central regions of the average $\Sigma_{\ast}$ (normalized to $h_{\rm R}$). 
}
\label{diskstack}
\end{center}
\end{figure}
%
%-------------------------------------------------------------
%
We use the 3.6~$\mu$m photometry from the Spitzer Survey of Stellar Structure (S$^{4}$G; \cite[Sheth et al. 2010]{Sheth2010}) for 1154 disk galaxies ($i<65^{\circ}$). 
\cite[D\'iaz-Garc\'ia et al. (2016)]{DG2016} de-projected and Fourier decomposed the images in a polar grid using the NIR-QB code (\cite[Salo et al. 1999]{Salo1999}; \cite[Laurikainen $\&$ Salo 2002]{Laurikainen2002}). 
Here, we resize the $m=0$ radial amplitudes ($I_{0}$) to a common frame defined by (i) the extent of the disks in physical units, and (ii) the disk scalelength ($h_{\rm R}$, from \cite[Salo et al. 2015]{Salo2015}). 
We stack the rescaled $I_{0}$ profiles to obtain 1-D average stellar density profiles ($\Sigma_{\ast}$) in bins of $M_{\ast}$ (from \cite[Mu\~noz-Mateos et al. 2015]{Munoz2015}) and Hubble type (Fig.\,1). 
For all the bins, the resulting profiles show an exponential decay down to at least $\sim 10 M_{\odot} \rm pc^{-2}$. 
The disk central surface brightness and scalelength of the mean $\Sigma_{\ast}$ increase with increasing $M_{\ast}$. 
We show that central mass concentrations in massive systems ($M_{\ast} \gtrsim 10^{10}M_{\odot}$) are substantially larger than in less massive galaxies on average. 
For the faintest systems, the mean $\Sigma_{\ast}$ presents hardly any deviation from the exponential disk. 

We find observational evidence for bar-induced secular evolution of a representative sample of disk galaxies, which is independent of any decomposition technique. 
For $M_{\ast} \ge 10^{9}M_{\odot}$, there is a significant difference in the mean $\Sigma_{\ast}$ between barred and non-barred systems, 
which is clear both for early- ($T<5$) and late-type ($T\ge5$) systems (shown in Fig.\,1): 
(i) disks in barred galaxies present larger scalelengths and fainter extrapolated central surface brightnesses;  
(ii) the average $\Sigma_{\ast}$ of barred and non-barred galaxies intersect each other very close to the mean bar length; 
and (iii) the central mass concentration of barred galaxies is larger (by almost a factor~2 when $T\le5$) than in their non-barred counterparts.
%
%-------------------------------------------------------------
%
\section{2-D stacked stellar bars} 
%
%-------------------------------------------------------------
%
Roughly 2/3 of the galaxies in our sample are barred according to \cite[Buta et al. (2015)]{Buta2015}. 
Prior to obtaining 2-D bar stacks, the images of barred galaxies are reconstructed from the Fourier modes and rotated with respect to the bar major axis, so that the final bar position angle is zero. 
If the spiral arms wind anticlockwise in the sky, we perform a geometric reflection across the bar major axis to make them winding clockwise. 
Finally, we resize the reoriented image to a grid of radius 3 times the bar radius. 

In Fig.\,2 we show 2-D synthetic bars resulting from co-adding the rescaled images in bins of total stellar mass and galaxy family 
($\underline{\rm A}$B/AB/A$\underline{\rm B}$/B, from \cite[Buta et al. 2015]{Buta2015}). 
We show that bars are oval-shaped among lenticular galaxies, and look intrinsically narrower in early- and intermediate-type spirals ($0\le T < 5$). 
However, among early-types the bar shape is rounded off by bulges and barlenses (\cite[Laurikainen et al. 2011]{Laurikainen2011}) in the inner parts. 
We also show a clear dependence of the bar ellipticity on the galaxy family. 

\begin{figure}[t]
\begin{center}
% \begin{tabular}{c c}
\subfloat{
\includegraphics[width=4.7in]{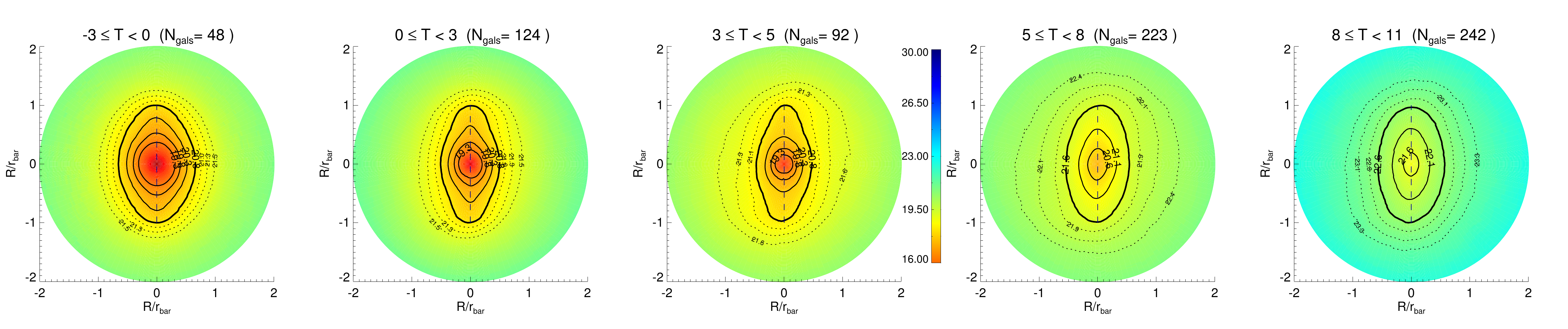}}\\[-2ex]
\vspace*{-0.5 cm}
\subfloat{
\includegraphics[width=3.8in]{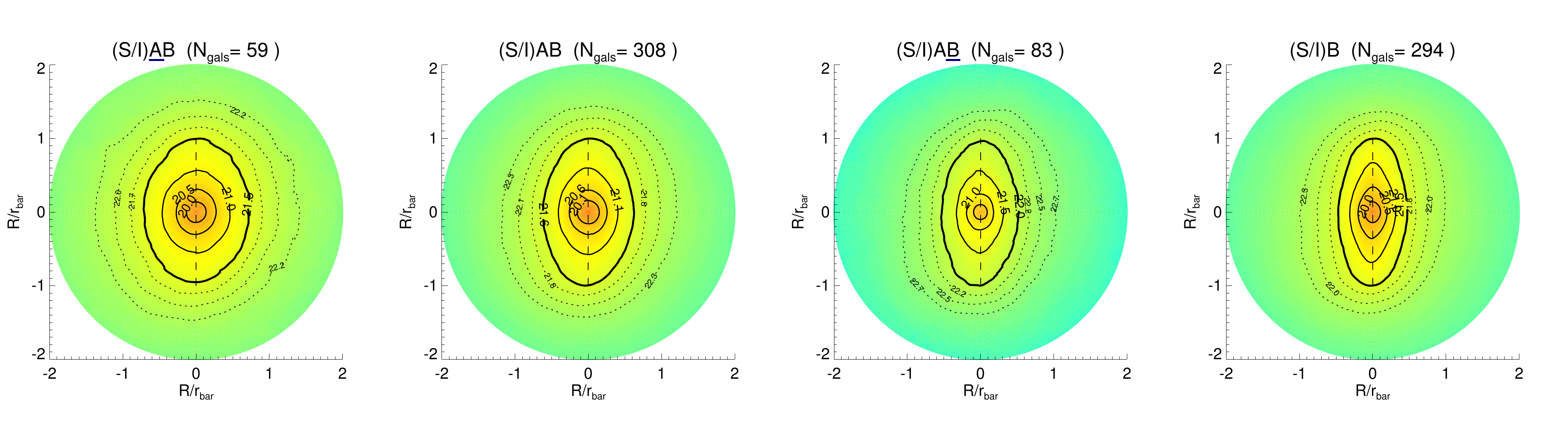}}
\label{stackbar}
% \end{tabular}
\end{center}
\vspace*{-0.3 cm}
\caption{
2-D bar stacks as a function of $M_{\ast}$ (upper row) and the galaxy family (lower row).
}
\end{figure}
%
%-------------------------------------------------------------
%
\begin{acknowledgements}
We acknowledge financial support to the DAGAL network from the People Programme (Marie Curie Actions)
of the European Union's Seventh Framework Programme FP7/2007- 2013/ under REA grant agreement number PITN-GA-2011-289313. 
%
% ########################################################################################### TRASH:
%
\end{acknowledgements}
%
%-------------------------------------------------------------
%
\vspace*{-0.6 cm}
%
%-------------------------------------------------------------
%


\begin{thebibliography}{}

\bibitem[Athanassoula (1992)]{Athanassoula1992}
{Athanassoula, E.} 1992,
\textit{MNRAS}, 259, 328

\bibitem[Athanassoula $\&$ Misiriotis (2002)]{Athanassoula2002}
{Athanassoula, E. $\&$ Misiriotis, A.} 2002,
\textit{MNRAS}, 330, 35

\bibitem[Athanassoula (2012)]{Athanassoula12}
{Athanassoula, E.} 2012,
\textit{MNRAS}, 426, L46

\bibitem[Athanassoula (2013)]{Athanassoula13}
{Athanassoula, E.} 2013,
\textit{Bars and secular evolution in disk galaxies: Theoretical input, ed. J. Falc\'on-Barroso $\&$ J. H. Knapen}, 305

\bibitem[Buta et al. (2015)]{Buta2015}
{Buta, R. J., Sheth, K., Athanassoula, E., et al.} 2015,
\textit{ApJS}, 217, 32

\bibitem[Debattista et al. (2006)]{Debattista06}
{Debattista, V. P., Mayer, L., Carollo, C. M. et al.} 2006,
\textit{ApJ}, 645, 209

\bibitem[D{\'{\i}}az-Garc{\'{\i}}a et al. (2016)]{DG2016}
{{D{\'{\i}}az-Garc{\'{\i}}a}, S., {Salo}, H., {Laurikainen}, E., $\&$ {Herrera-Endoqui}, M.} 2016,
\textit{A$\&$A}, 587, A160

\bibitem[Herrera-Endoqui et al. (2015)]{HE2015}
{{Herrera-Endoqui}, M., {D{\'{\i}}az-Garc{\'{\i}}a}, S., {Laurikainen}, E., $\&$ {Salo}, H.} 2015,
\textit{A$\&$A}, 582, A86

\bibitem[Hohl (1971)]{Hohl71}
{Hohl, F.} 1971,
\textit{ApJ}, 168, 343

\bibitem[Knapen et al. (2000)]{Knapen2000}
{Knapen, J. H., Shlosman, I., $\&$ Peletier, R. F.} 2000,
\textit{ApJ}, 529, 93

\bibitem[Laurikainen $\&$ Salo (2002)]{Laurikainen2002}
{Laurikainen, E. $\&$ Salo, H.} 2002,
\textit{MNRAS}, 337, 1118

\bibitem[Laurikainen et al. (2004)]{Laurikainen2004}
{Laurikainen, E., Salo, H., $\&$ Buta, R.} 2004,
\textit{ApJ}, 607, 103

\bibitem[Laurikainen et al. (2011)]{Laurikainen2011}
{Laurikainen, E., Salo, H., Buta, R., $\&$ Knapen, J. H.} 2011,
\textit{MNRAS}, 418, 1452

\bibitem[Minchev et al. (2011)]{Minchev11}
{Minchev, I., Famaey, B., Combes, F.,} 2011,
\textit{A$\&$A}, 527, A147

\bibitem[Mu\~noz-Mateos et al. (2015)]{Munoz2015}
{Mu\~noz-Mateos, J. C., Sheth, K., Regan, M., et al.} 2015,
\textit{ApJS}, 219, 3

\bibitem[Salo et al. (1999)]{Salo1999}
{Salo, H., Rautiainen, P., Buta, R., et al.} 1999,
\textit{AJ}, 117, 792

\bibitem[Salo et al. (2015)]{Salo2015}
{Salo, H., Laurikainen, E., Laine, J., et al.} 2015,
\textit{ApJS}, 219, 4

\bibitem[S{\'a}nchez-Janssen $\&$ Gadotti (2013)]{SG2013}
{{S{\'a}nchez-Janssen}, R. $\&$ {Gadotti}, D.~A.} 2013,
\textit{MNRAS}, 432, L56

\bibitem[Sheth et al. (2010)]{Sheth2010}
{Sheth, K., Regan, M., Hinz, J. L., et al.} 2010,
\textit{PASP}, 122, 1397

\bibitem[Wada $\&$ Habe (1992)]{Wada1992}
{Wada, K. $\&$ Habe, A.} 1992,
\textit{MNRAS}, 258, 82

\end{thebibliography}
\end{document}